\documentclass[preprint]{aastex}
\usepackage{natbib}
\usepackage{booktabs}
\usepackage{threeparttable}
\usepackage{longtable}
\usepackage{tabularx}
\usepackage{lineno}

\newcommand \raa {Research in Astronomy and Astrophysics}

\begin{document}

\title{Very long-periodic pulsations detected simultaneously in a white-light flare and sunspot penumbra}

\author{Dong~Li$^{1,2}$, Jincheng~Wang$^{3,2}$, and Yu~Huang$^1$}
\affil{$^1$Key Laboratory of Dark Matter and Space Astronomy, Purple Mountain Observatory, CAS, Nanjing 210023, PR China \\
       $^2$Yunnan Key Laboratory of the Solar physics and Space Science, Kunming 650216, PR China \\
       $^3$Yunnan Observatories, Chinese Academy of Sciences, Kunming 650216, China}
     \altaffiltext{}{Correspondence should be sent to: lidong@pmo.ac.cn}
\begin{abstract}
We investigate the origin of very long-periodic pulsations (VLPs) in
the white-light emission of an X6.4 flare on 2024 February 22
(SOL2024-02-22T22:08), which occurred at the edge of a sunspot
group. The flare white-light fluxes reveal four successive and
repetitive pulsations, which are simultaneously measured by the
Helioseismic and Magnetic Imager and the White-light Solar
Telescope. A quasi-period of 8.6$^{+1.5}_{-1.9}$~minutes, determined
by the Morlet wavelet transform, is detected in the visible
continuum channel. The modulation depth, which is defined as the
ratio between the oscillatory amplitude and its long-term trend, is
smaller than 0.1\%, implying that the QPP feature is a weak
wave process. Imaging observations show that the X6.4 flare occurs
near a sunspot group. Moreover, the white-light brightening is located
in sunspot penumbra, and a similar quasi-period of about
8.5$^{+1.6}_{-1.8}$~minutes is identified in one penumbral location
of the nearest sunspot. The map of Fourier power
distribution suggests that a similar periodicity is universally
existing in most parts of the penumbra that is close to the
penumbral-photospheric boundary. Our observations support the
scenario of that the white-light QPP is probably modulated by the
slow-mode magnetoacoustic gravity wave leaking from the sunspot
penumbra.
\end{abstract}


\section{Introduction}
The flare seen in the visible continuum is often called the white-light
flare (i.e., WLF), which was first reported by Carrington
\citep{Carrington59}. It is a rare flare event compared to the soft
X-ray (SXR) flare on the Sun, mainly because of the strong optical
background in the photosphere; for instance, the enhancement of the
visible continuum during the WLF is typically 5\%$-$50\%
\citep{Jess08,Namekata17,Liy24}. Based on the flare location in the
solar surface, WLFs are classified into off-limb and on-disk flares.
The off-limb WLF, with a visible continuum source detected beyond the
solar limb, is only reported in a few cases
\citep{Oliveros14,Fremstad23}. Conversely, the on-disk optical
continuum brightening has been reported in the majority of the WLFs
\citep{Song18,Joshi21}. It is widely known that the visible
continuum emissions of WLFs are strongly correlated with the
microwave and hard X-ray (HXR) emissions temporally and spatially
\citep{Krucker15,Li23}. Such a strong correlation may indicate that
the flare white-light radiation is highly associated with electron
beams that are accelerated by the flare energy \citep{Ding03}.
However, those nonthermal electrons hardly penetrate into the
photosphere, where the originated source of optical continua is located
\citep{Najita70}. Thus, the white-light continuum emission may not
be fully heated by a pure mechanism of electron beams, and some
portions of the white-light enhancement might be caused by an
additional source, such as the Alfv\'{e}n wave or the hydrogen
free-free emission \citep{Heinzel17,Song23}. This type of WLF
appears to be easily seen in powerful flares on the Sun
\citep{Zhao21,Li23b}, and they are also frequently observed in
superflares in solar-type stars \citep{Shibayama13,Yan21} since
they both release a large amount of energies.

One common feature that is associated with the flare radiation is
quasi-periodic pulsations (QPPs) in the time series during
solar/stellar flares. A typical QPP is generally characterized by at
least three successive and repeated pulsations in the time-dependent
intensity curves \citep{Zimovets21}. The time duration of each
pulsation for one flare QPP is expected to be stationary, which is
regarded as the period. However, the time scales for the observed
pulsations are always non-stationary, termed as quasi-periods of
non-stationarity QPPs \citep{Nakariakov19}. The quasi-periods can be
detected in a wide timescale ranging from subseconds through dozens
of seconds to a few minutes
\citep{Tan10,Hayes20,Kashapova20,Li20a,Li21,Li23c,Shen22,Collier23,Karlicky23}.
The flare QPPs on the Sun are frequently observed in a broad
wavelength range of radio/microwave, H$\alpha$, ultraviolet (UV),
extreme ultraviolet (EUV), SXR/HXR, and even
$\gamma$-rays
\citep{Nakariakov10,Dolla12,Tan16,Dominique18,Li22a,Lorincik22,Zimovets23,Millar24,Zhou24}.
However, they are rarely reported in the white-light continuum
emission during solar flares, which may because WLFs are difficult
to be detected \citep{Zhao21}. On the other hand, the flare QPPs in
white-light wave bands are often observed in solar-type stars, and
their quasi-periods are always ultralong, i.e., $>$10~minutes
\citep{Pugh16,Kolotkov21}, although some short-period QPPs are seen
in stellar flares \citep{Howard22}. It appears that the reported
periods are strongly dependent on the time resolution of the
observed instruments, including the QPPs both in the solar and stellar
flare.

The generation mechanism of flare QPPs is still under debate
\citep{Zimovets21}. Generally, the flare QPPs are directly modulated
by the eigenmodes of magnetohydrodynamic (MHD) waves in magnetic
loops or current sheets, such as the fast kink and sausage modes,
the torsional Alfv\'{e}n mode, and the slow mode
\citep{Nakariakov20}. The flare QPPs can also be triggered by a
quasi-periodic regime of magnetic reconnection, and the
quasi-periodic reconnection could be either spontaneous or induced
by external waves \citep{Takasao16,Karampelas23}. The idea is that
the nonthermal electrons are periodically accelerated by the
repetitive reconnection during solar flares, and thus it is easier
to cause the flare QPP that was observed simultaneously in HXR and
microwave emissions during the impulsive phase \citep{Yuan19,Li22b}.
It seems that the flare QPPs in different categories are driven by
various mechanisms, and one single mechanism cannot fully explain
all existing QPPs, which may because the available observations cannot
be adequate to distinguish between different mechanisms \citep{Inglis23}.

One critical issue is to understand why QPPs in the white-light
continuum are rarely observed in solar flares \citep{Zhao21} but
are frequently found in white-light fluxes of stellar flares
\citep{Pugh16,Kolotkov21,Howard22}; another is to unlock the
generation mechanism of flare QPPs in the optical continuum
\citep{Zimovets21}. In this Letter, we investigated the flare QPP
with a very long quasi-period in the white-light continuum and also
attempted to reveal its driver. The Letter is organized as follows:
Section~2 introduces observations and instruments, Section~3
presents the data analysis and main results, Section~4 offers some
discussions, and a brief summary is given in Section~5.

\section{Observations and Instruments}
We analyzed the solar flare occurred in the active region of
NOAA~13234 on 2024 February 22, which is located near a sunspot group.
It was simultaneously measured by the space- and ground-based
telescopes, that is, the Full-disk Vector MagnetoGraph
\citep[FMG;][]{Deng19}, the Ly$\alpha$ Solar Telescope
\citep[LST;][]{Feng19}, and the Hard X-ray Imager
\citep[HXI;][]{Suy19} on aboard the Advanced Space-based Solar
Observatory (ASO-S), the Atmospheric Imaging Assembly
\citep[AIA;][]{Lemen12} and the Helioseismic and Magnetic Imager
\citep[HMI;][]{Schou12} on board the Solar Dynamics Observatory
(SDO), the Geostationary Operational Environmental Satellite (GOES),
the Nobeyama Radio Polarimeters (NoRP), and the Expanded Owens
Valley Solar Array (EOVSA).

ASO-S is a new space observatory that explores the connection
between solar magnetic fields, solar flares, and coronal mass
ejections, which has three payloads on board. FMG is designed to
measure the full-disk photospheric magnetic field, and now it
provides the local line-of-sight (LOS) magnetogram in the active
region and also provides the white-light image at the same active
region in the optical continuum channel near the Fe 5324~{\AA} line.
They have the same temporal cadence of about 120~s and a spatial
resolution of $\sim$1.5$^{\prime\prime}$. LST is composed of two
telescopes that see the entire solar disk, the Solar Disk Imager
(SDI) captures the Ly$\alpha$ snapshot at 1216~{\AA}, and the
White-light Solar Telescope (WST) takes the white-light map at
3600~{\AA}. The white-light map measured by WST has a temporal
cadence of 120~s in the regular mode and changes to 1~s or 2~s in
the flare mode. HXI takes the flare imaging spectroscopy in the HXR
energy range of $\sim$10$-$300~keV. The temporal cadence is 4~s in
the regular mode and can reach to be 0.125~s in the burst mode. In
this work, we used the LOS magnetogram measured by FMG, the WST map
at 3600~{\AA}, the HXI light curves, and a reconstructed map in the
energy range of 20$-$50~keV.

SDO/AIA takes full-disk solar maps at multiple EUV/UV wave bands
nearly simultaneously. The AIA maps in wave bands of UV~1600~{\AA}
and EUV~131~{\AA} are analyzed, and their temporal cadences are 24~s
and 12~s, respectively. SDO/HMI provides the full-disk solar
magnetogram, dopplergram, and continuum filtergram. We analyzed the
visible continuum images near the Fe~6173~{\AA} line, which has a
time cadence of 45~s. Both AIA and HMI maps were preprocessed by
the standard procedures of ``aia\_prep.pro'' and ``hmi\_prep.pro'',
and they have a same spatial resolution of 1.2$^{\prime\prime}$. The
HMI vector magnetogram is also used to analyze the magnetic topology
in the flare area, which has a spatial resolution of
1.0$^{\prime\prime}$ and a time cadence of 720~s. NoRP records
full-disk solar fluxes at six microwave frequencies with a temporal
cadence of 1~s. EOVSA provides the solar radio dynamic spectrum in
the microwave frequency range of $\sim$1$-$18~GHz with a temporal
cadence of about 1~s, and it can also measure the solar maps in the
microwave channels. However, only a few maps are provided
online\footnote{http://ovsa.njit.edu/fits/flares/2024/02/22/20240222223200/},
and one EOVSA map was used to show the flare profile in the
microwave channel.

\section{Data analysis and Results}
\begin{figure}
\centering
\includegraphics[width=\linewidth,clip=]{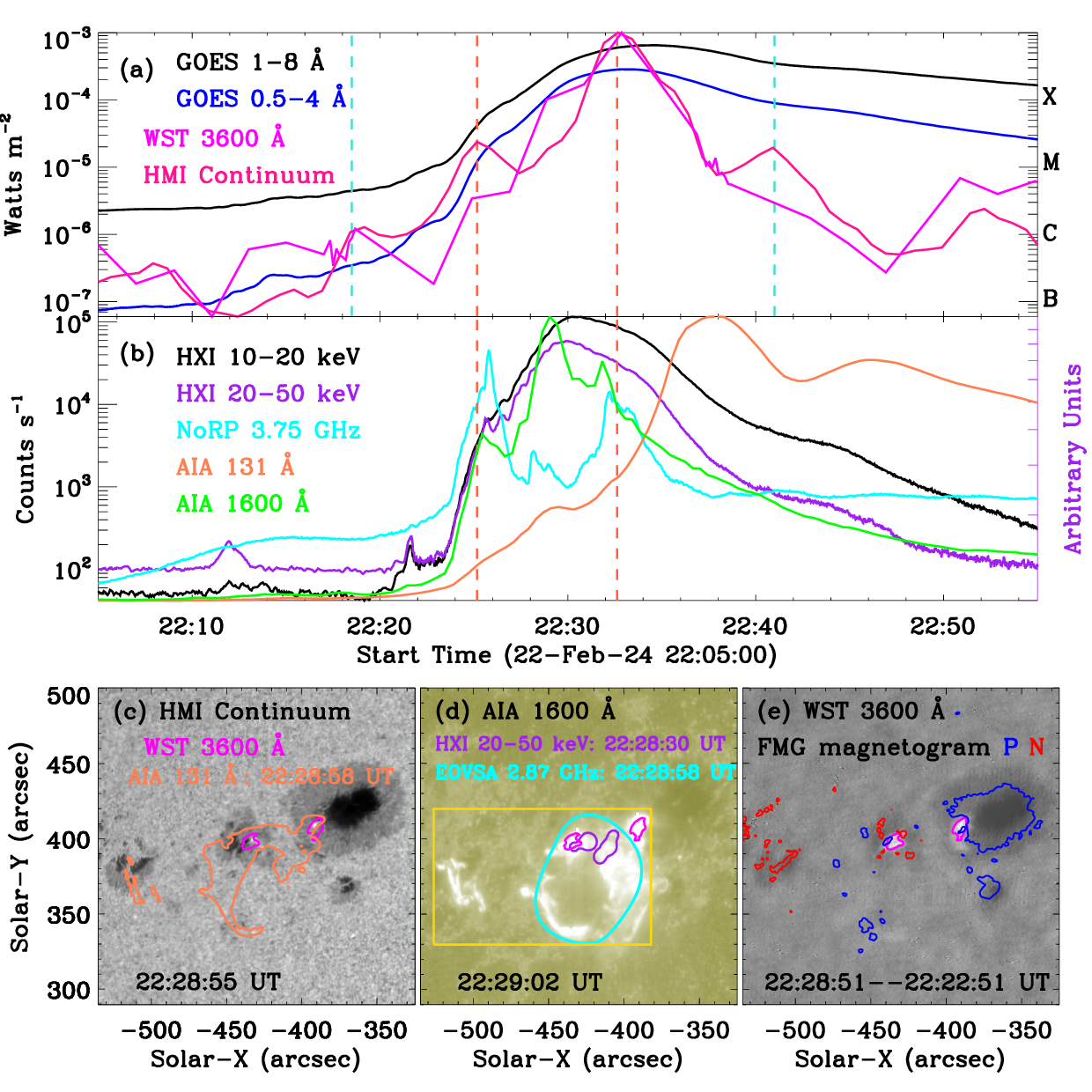}
\caption{Overview of the solar flare on 2024 February 22. (a): SXR
light curves recorded by GOES at 1$-$8~{\AA} (black) and
0.5$-$4~{\AA} (blue). White-light fluxes at WST~3600~{\AA}
(magenta), and HMI continuum (deep pink). The vertical lines
(turquoise and tomato) mark four white-light peaks. (b): HXR and
microwave fluxes at HXI~10$-$20~keV (black) and 20$-$50~keV
(purple), and NoRP~3.75~GHz (cyan), as well as the EUV/UV fluxes at
AIA~131~{\AA} (coral) and 1600~{\AA} (green). (c): The
pseudo-intensity map with a FOV of
$\sim$210\arcsec$\times$210\arcsec\ derived from the HMI continuum
data. The coral contour represents the flare radiation at
AIA~131~{\AA}. (d): UV map with the same FOV at AIA~1600~{\AA}. The
purple and cyan contours outline the HXR and microwave sources at
HXI~20$-$50~keV and EOVSA~2.87~GHz, respectively. The levels are set
at 15\% of their maximum. The gold rectangle marks the flare area.
(e): Base difference map derived from the WST~3600~{\AA} data. The
blue and red contours represent the positive (P) and negative (N)
magnetic fields at levels of $\pm$1000~G measured by FMG. The
magenta contour outlines the flare radiation at WST~3600~{\AA}.
\label{over}}
\end{figure}

Figure~\ref{over}~(a) presents the light curves in SXR and
white-light channels from 22:05~UT to 22:55~UT on 2024 February 22.
The GOES flux in the SXR channel of 1$-$8~{\AA} (black) suggests an
X6.4-class flare, and it begins at 22:08~UT, peaks at 22:34~UT, and
stops at
22:43~UT\footnote{https://www.solarmonitor.org/?date=20240223}. The
GOES flux at 0.5$-$4~{\AA} (blue) reveals a similar profile to
that at 1$-$8~{\AA}. Conversely, the white-light intensity curve
measured by HMI in the visible continuum channel (deep pink)
exhibits at least four successive pulsations during the flare, as
marked by the four vertical dashed lines. Those successive
pulsations have an average duration of roughly 8~minutes, which may
be regarded as the flare QPP in the white-light emission. There are
some successive peaks in the white-light time series (magenta)
measured by WST at 3600~{\AA}, and they appear to be consistent with
those pulsations seen in the visible continuum channel, confirming
the presence of white-light QPPs. We note that the time series at
WST~3600~{\AA} show some burst noises at about 22:18~UT and
22:37~UT, which may be due to the switching of the regular and burst
modes, resulting in various temporal resolutions of WST.
Figure~\ref{over}~(b) shows the light curves in wavelengths of HXR,
microwave, and EUV/UV during 22:05$-$22:55~UT. We cannot see the
four successive pulsations in these light curves, which are
different from the white-light flux profile. The HXR fluxes recorded
by HXI (black and purple) appear as a series of successive pulses at a
short time scale, and their amplitudes are very small. The light
curves in wave bands of microwave (cyan) and UV flux (green) seem to
reveal three successive pulses with a large amplitude, but their
time duration is short \citep{Lid24}. Two pulses in microwave and UV
wave bands appear to match the pulsations in white-light channels,
although they are not strictly the one-to-one correspondence, as
indicated by two tomato vertical lines. The EUV flux at
AIA~131~{\AA} (coral) also reveals three successive pulsations, but
they are later than the white-light pulsations, and their duration
is much longer. Here, to avoid those saturated images, the time
cadence of AIA~131~{\AA} is chosen to be 24~s. Briefly, the
white-light QPP cannot be seen in wave bands of HXR, microwave, and
EUV/UV. At last, we want to state that the SXR/HXR and microwave
fluxes recorded by GOES, HXI, and NoRP are all integrated over the
entire Sun, while the time series in channels of white-light and
EUV/UV are integrated over the flare area, dubbed the local flux.

In Figure~\ref{over}~(c)$-$(e), we show multi-wavelength images
during the X6.4 flare. Panel~(c) plots the pseudo-intensity map with
a field-view-of (FOV) of
$\sim$210$^{\prime\prime}$$\times$210$^{\prime\prime}$, which is
derived from the HMI continuum images near the Fe~6173~{\AA} line.
Here, the white-light emission is significantly enhanced in the
pseudo-intensity map \citep{Song18b}. It can be seen that there are two
patches of visible continuum enhancements in the pseudo-intensity
map, and they are located at the edge of two sunspot penumbras.
Double bright kernels are also seen in the white-light
image captured by WST at 3600~{\AA}. They both match two patches in
the visible continuum channel, as indicated by the two magenta contours.
These two patches appear to be connected by a hot loop system seen
at AIA~131~{\AA}, as shown by the coral contours.
Figure~\ref{over}~(d) draws the UV map at AIA~1600~{\AA}, which
reveals a main quasi-circular structure and a remote ribbon-like
shape. This observation indicates that the X6.4 flare could be a
circular-ribbon flare. The gold rectangle marks the flare area used
to integrate the local flux in wave bands of white light and EUV/UV.
The microwave emission measured by EOVSA at 2.87~GHz displays a
ellipse profile shape (cyan contour), and it matches the main
quasi-circular structure, further confirming the presence of a
circular-ribbon flare. Two HXR sources appear in the energy range of
HXI~20$-$50~keV, which are overlaid on the main quasi-circular
profile, as indicated by the purple contours. Here, the HXR map is
reconstructed by the HXI\_CLEAN algorithm, utilizing the detectors
from D29 to D91. The fine grids of G1 to G3 are excluded since they
are not yet calibrated well \citep{Li23c}. We can find that one HXR
source appears to match a white-light patch, but the other HXR source
is far away from another white-light patch. That is, the two HXR
sources are not consistent with the double white-light patches.
Figure~\ref{over}~(e) shows the difference map observed by WST at
3600~{\AA}, which clearly reveals two bright kernels. Here, the base
difference map is plotted, so the white-light emissions can be
enhanced. The blue and red contours represent the positive and
negative magnetic fields measured by FMG, and the main
quasi-circular structure of the X6.4 flare is mainly surrounded by
positive magnetic fields.

\begin{figure}
\centering
\includegraphics[width=\linewidth,clip=]{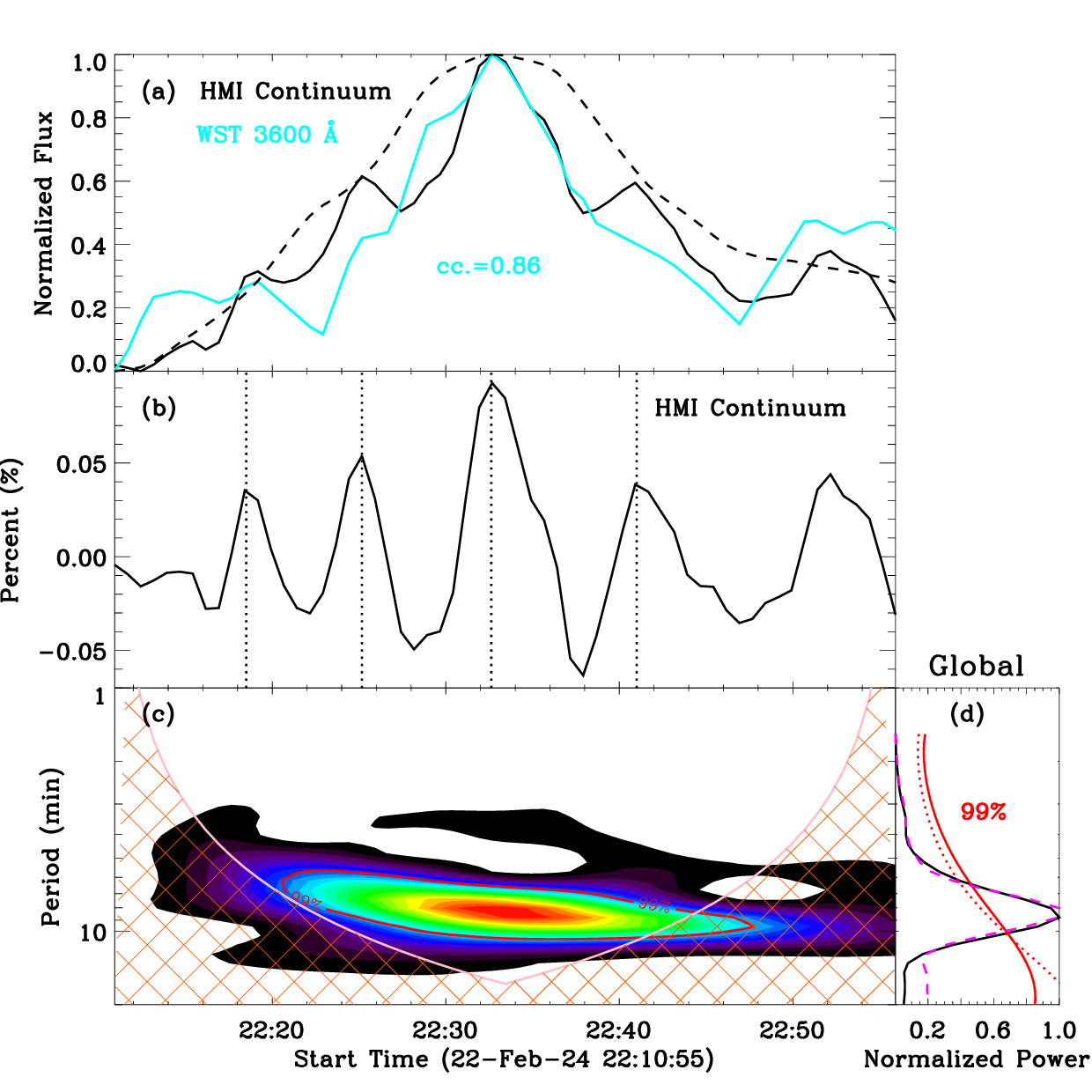}
\caption{Morlet wavelet analysis results. (a): White-light fluxes
measured by the HMI continuum (black) and WST~3600~{\AA} (cyan). The
overlaid dashed line represents the long-term trend at the HMI
continuum. Here, the WST~3600~{\AA} flux has been interpolated into
a uniform time cadence of 45~s. (b): Detrended light curve
normalized to its long-term trend at the HMI continuum. (c): Morlet
wavelet power spectrum. (d): Global wavelet power spectra for the
running windows of 10~minutes (black) and 15~minutes (magenta),
respectively. The red contour and lines represent a significance
level of 99\%. \label{wav1}}
\end{figure}

In order to identify the quasi-period of the flare QPP, the wavelet
transform with a mother function \citep{Torrence98} is performed for
the detrended time series in the visible continuum channel. The
detrended time series is accomplished by subtracting the 10 minute
running average from its raw intensity curve, and thus the long-term
trend is suppressed, while the short-period QPP can be strengthened
\citep{Yuan11,Tian12}. Figure~\ref{wav1} presents the Morlet wavelet
analysis results for white-light fluxes. Panel~(a) shows the raw
white-light fluxes in the HMI continuum (black) and WST~3600~{\AA}
(cyan) channels, as well as the long-term trend at the HMI continuum
(dashed curve). Here, the raw flux has been used as normalization,
i.e., $\frac{f-f_{\rm min}}{f_{\rm max}-f_{\rm min}}$, where $f$ is
the observed flux and $f_{\rm min}$ and $f_{\rm max}$ represent the
minimum and maximum fluxes, respectively. The white-light flux at
WST~3600~{\AA} has been interpolated as a uniform cadence of 45~s,
which is the same as that at the HMI continuum. Then, the linear Pearson
correlation coefficient (cc.) of the two white-light fluxes is
estimated to about 0.86, implying a high correlation between them.
Therefore, only the Morlet wavelet analysis results at the HMI continuum
are shown. Panel~(b) plots the normalized detrended time series in
the HMI continuum channel as normalization by its long-term trend.
It clearly shows four successive pulsations (vertical lines), which
match those in the raw light curve, confirming that the running average
can only enhance the QPP feature but not change the period. The
modulation depth, which is determined by the ratio between the
oscillatory amplitude and its long-term trend, is about
0.05\%$-$0.09\%. Panels~(c) and (d) draw the Morlet wavelet power
spectrum and its global wavelet power spectrum. They are both
dominated by a bulk of the power spectrum inside the 99\% significance
level, suggesting a dominant period with a large uncertainty. The
dominant period is determined by the peak of the global wavelet
power spectrum, and its uncertainty is estimated from the full width
at the half of the 99\% significance level, which is about
8.6$^{+1.5}_{-1.9}$~minutes. We note that the 10-min running
window is very close to the measured period, which might be caused
by the artifact of the detrending process. Therefore, a long running
window of 15~minutes is used for detrending, and the global wavelet
power spectrum is shown in Figure~\ref{wav1}~(d). The same
quasi-period of about 8.6$^{+1.5}_{-1.9}$~minutes is identified with
the same method, as indicated by the magenta dashed curve and red
dotted line. Moreover, the measured period is consistent with the
average duration of the four white-light pulsations, suggesting that
the white-light QPP is not a false periodicity.

\begin{figure}
\centering
\includegraphics[width=\linewidth,clip=]{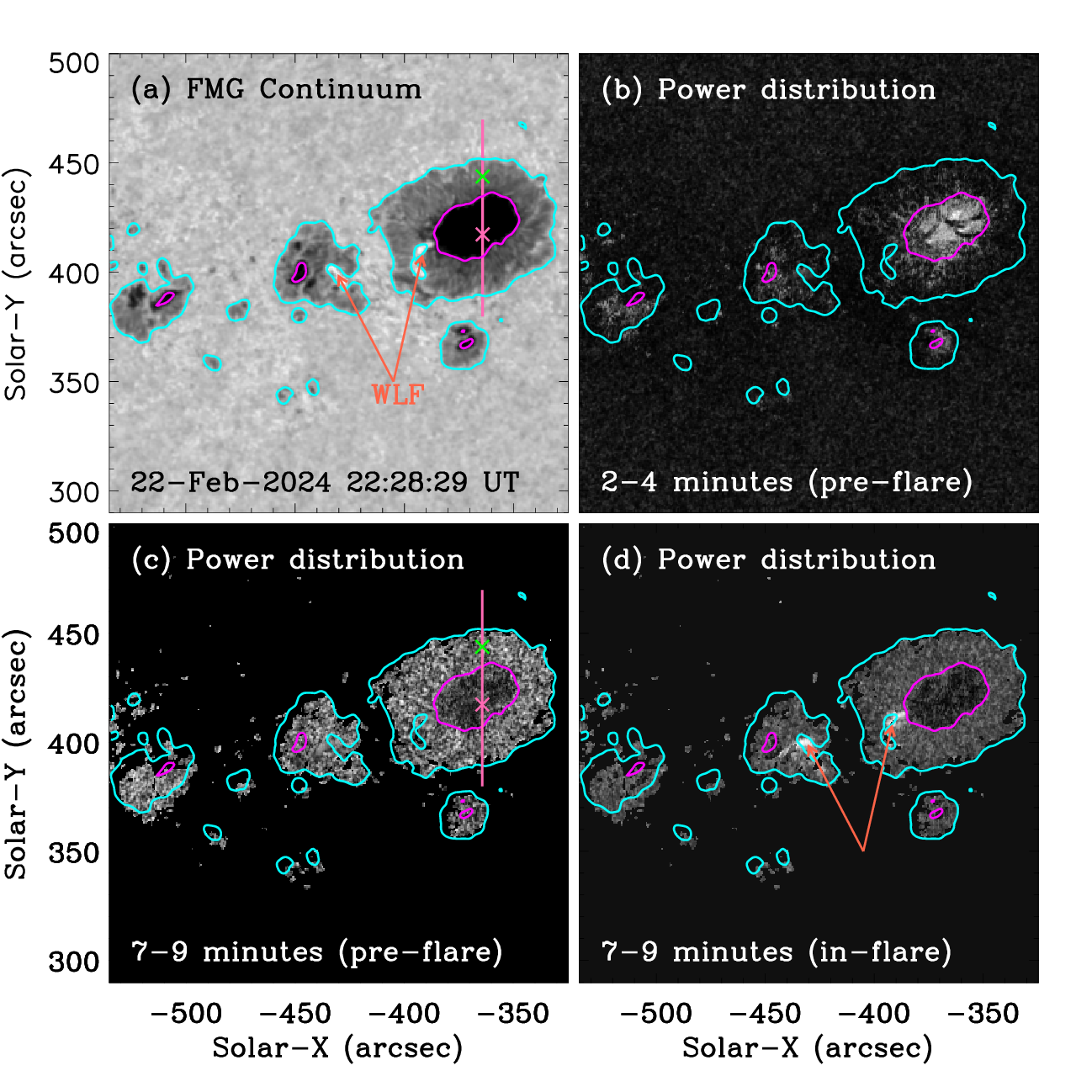}
\caption{(a) The optical continuum image observed by ASO-S/FMG. The
overplotted contours outline the penumbral-photospheric (cyan) and
umbral-penumbral (magenta) boundaries, which are derived from the
optical continuum radiation in the Fe 5324~{\AA} line. (b) \& (c)
Fourier power maps that are averaged over 2-4~minutes and
7-9~minutes before the X6.4 flare. The hot pink line outline the
slit position crosses the sunspot. The pluses ($`+'$) mark the
positions used to integrate the intensity curves in the penumbra
(green) and umbra (hot pink). (d): Fourier power map
that is averaged over 7-9~minutes during the WLF. The tomato arrows
indicate the white-light brightening. \label{pow}}
\end{figure}

In order to search for the trigger source of the white-light QPP, we
perform the Fourier transform \citep{Inglis08,Yuan19} on the
emission intensity of every pixel in the HMI continuum channel, as
shown in Figure~\ref{pow}. Panel~(a) presents the optical continuum
map measured by ASO-S/FMG near the Fe~5324~{\AA} line, which clearly
shows two groups of bright features coinciding with the two bright
kernels seen in the WST~3600~{\AA} and HMI continuum channels, as
indicated by the tomato arrows. The color contours are derived from
the optical continuum radiation, representing the
penumbral-photospheric (cyan) and umbral-penumbral (magenta)
boundaries. The hot pink line outlines a slit section
that crosses the central part of the sunspot, including the umbra
(i.e, magenta plus) and penumbra (i.e, green plus).
Figures~\ref{pow}~(b) \& (c) show spatial distributions of the
normalized Fourier power, and they are averaged among the spectral
components of 2$-$4~minutes and 7$-$9~minutes before the X6.4 flare.
We note that the short-period component at 2$-$4~minutes is mainly
located in the umbral region, and the long-period component at
7$-$9~minutes tends to appear in the bulk of the penumbra,
especially in the surrounding area that is close to the
penumbral-photospheric boundary. It is known that the
Fourier power will vary with time. To trace the sources of these
variations in time with the flare localization, panel~(d) presents
the spatial distribution of the normalized Fourier power, which is
averaged among the spectral components at 7$-$9~minutes during the
WLF. One can immediately note that the long-period
component is situated in the penumbral region. Moreover, the
long-period component is especially obvious in the WLF
site, as marked by the tomato arrows. This is sufficient to fix a
possible link between the oscillations in the sunspot penumbra and the
origin of the energy release in the WLF. Therefore,
the white-light QPP is most likely to be related to the long-period
component in the penumbra.

\begin{figure}
\centering
\includegraphics[width=\linewidth,clip=]{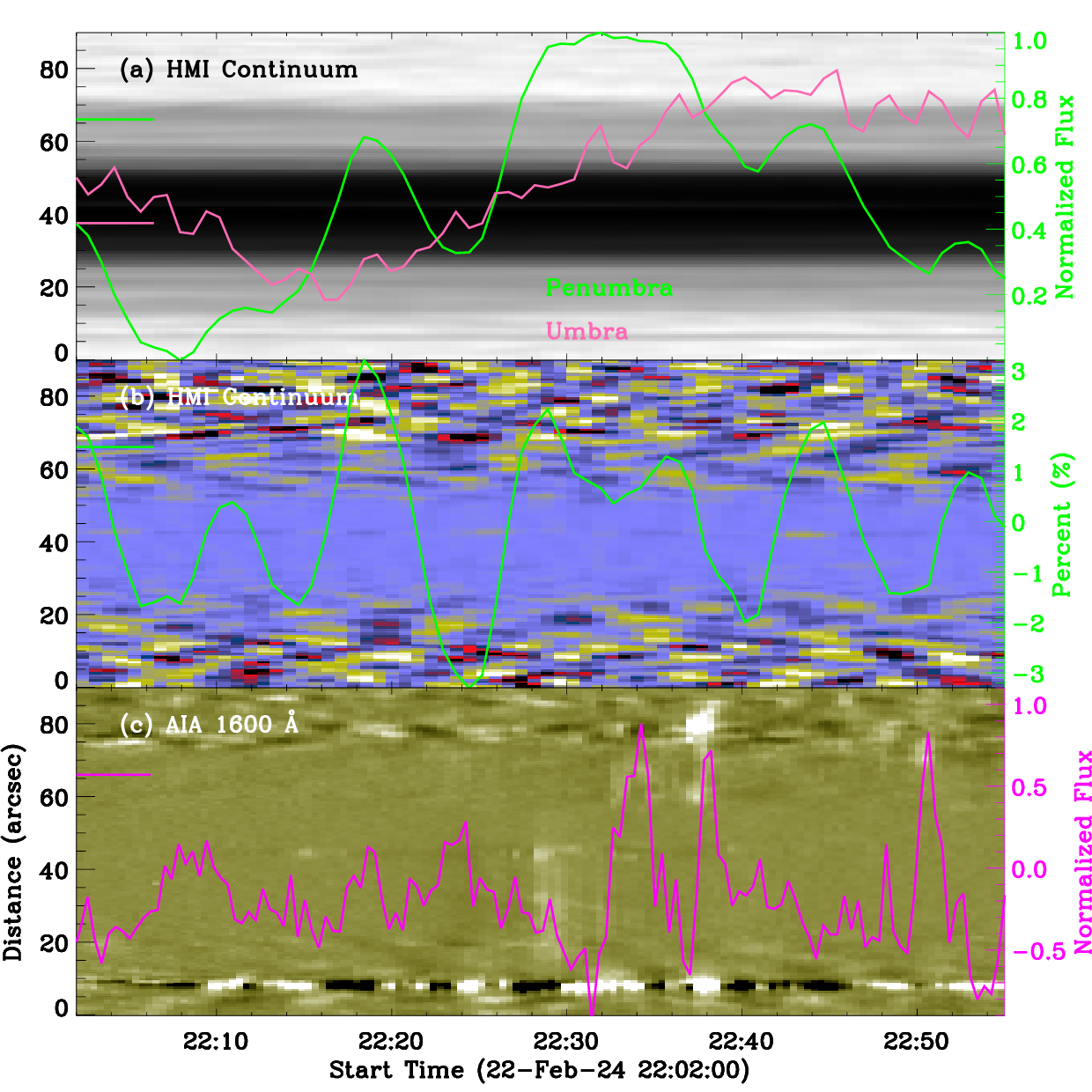}
\caption{Time-distance maps along the slit (hot pink line in
Figure~\ref{over}) that crosses the sunspot. (a): Time-distance map
at HMI continuum. (b) \& (c): Time-distance maps after removing the
10-minutes running average at the HMI continuum and AIA~1600~{\AA}. The
overplotted curves are the time series that extracted at the
penumbral and umbral positions, which are marked by the short lines
on the left.  \label{slit}}
\end{figure}

To look closely at the trigger source of the white-light QPP, we then
prepare the time-distance (TD) map along a cut slit that crosses the
sunspot umbra and penumbra, and the slit position is marked by a hot
pink line in Figure~\ref{pow}~(a). The selected slit is far
from the flare site, because we wanted to state that the periodicity
in the sunspot is long-standing but not affected by the flare
QPP. Figure~\ref{slit} shows the TD maps derived from the data cube
observed by HMI and AIA in the visible continuum and UV~1600~{\AA}.
Panel~(a) presents the raw TD map in the HMI continuum channel,
which can easily distinguish the penumbral and umbral regions, as
indicated by the short green and magenta lines on the left.
Then, the normalized intensity curves are extracted from the
penumbral and umbral positions, as indicated by the green and
magenta pluses in Figure~\ref{pow}. Note that the intensity curves
are both averaged over 5~pixels, so that the signal-to-noise ratio
is improved. We can see that the intensity curve at the penumbra
reveals several large pulsations with a long time scale, while that
at the umbra exhibits a number of small pulses with a short time
duration. Figure~\ref{slit}~(b) shows the detrended TD map after
removing the 10-minutes running average \citep{Yuan11,Tian12} in the
HMI continuum channel, which clearly shows the large pulsations at
the penumbra position, where is close to the penumbral-photospheric
boundary. The overplotted curve is the detrended intensity curve
normalized to its long-term trend, which also shows those large
pulsations with a long time scale. We note that the modulation depth
is about 2\%$-$3\%, which is much bigger than that of the
white-light QPP. Panel~(c) plots the detrended TD map in
AIA~1600~{\AA}, and we cannot see any apparent signature of QPPs at
the penumbral and umbral regions. Here, the normalized detrended
intensity curve at the penumbra shows some small wiggles, but they
are absolutely different from those large pulsations in the HMI
continuum channel, as indicated by the overplotted curve. All these
observations suggest that the long-periodic pulsations can be seen
at the sunspot penumbra in the photosphere, and they cannot propagate
upwardly into the chromosphere and transition region.

\begin{figure}
\centering
\includegraphics[width=\linewidth,clip=]{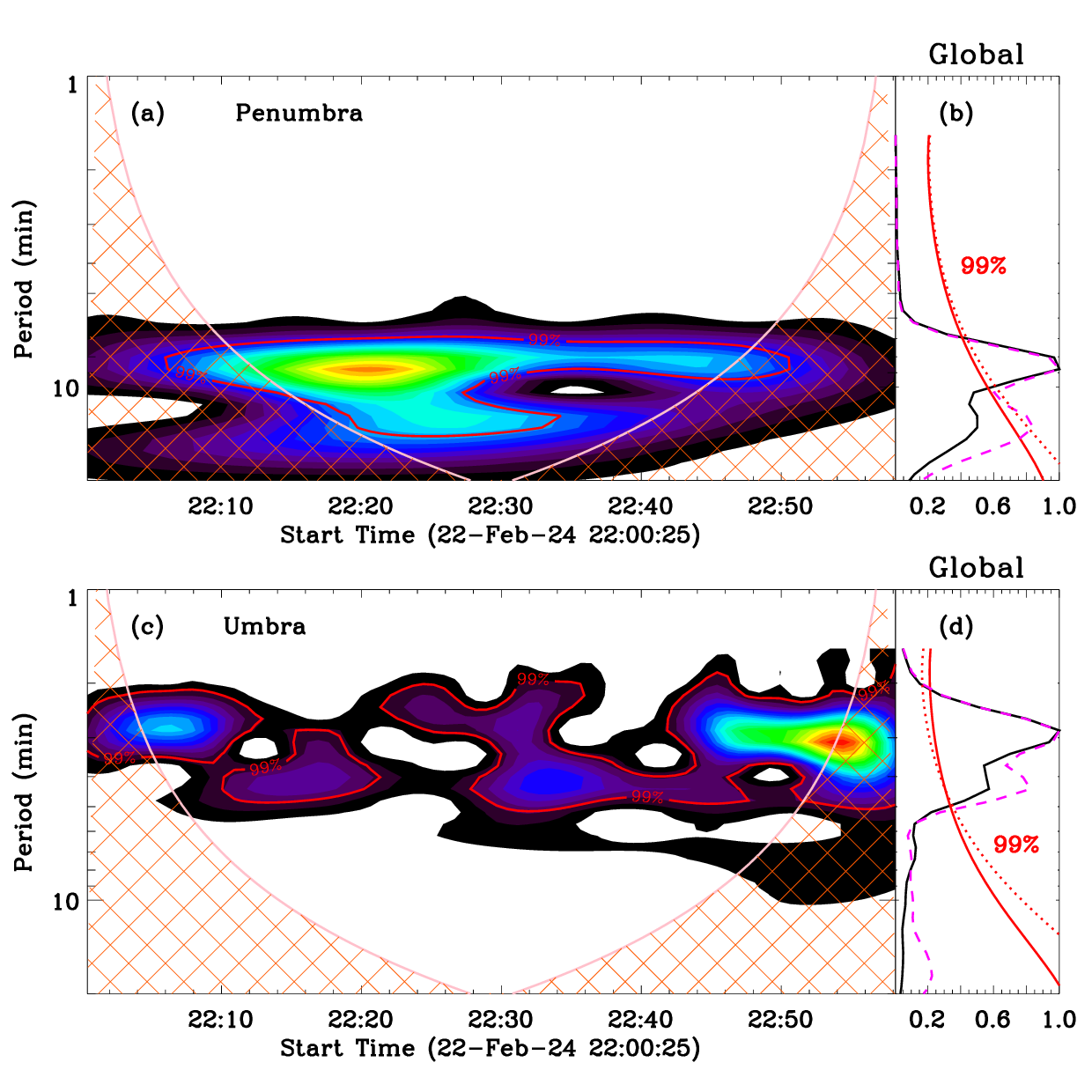}
\caption{Morlet wavelet analysis results for the intensity curves at
the penumbra (a \& b) and umbra (c \& d) of the adjacent sunspot.
The black and magenta curves in panels~(b) and (d) show the global
wavelet power spectra for the running windows of 10~minutes and
15~minutes, respectively. \label{wav2}}
\end{figure}

In order to determine the quasi-periods at the sunspot penumbra and
umbra, we also perform the same wavelet transform for the detrended
intensity curves in the visible continuum channel, as shown in
Figure~\ref{wav2}. Panels~(a) and (b) present the Morlet wavelet
power spectrum and its global wavelet power spectrum at the sunspot
penumbra. They are characterized by a bulk of the power spectrum
inside the 99\% significance level, and a dominant period within a
large uncertainty is estimated to about 8.5$^{+1.6}_{-1.8}$~minutes.
This quasi-period is exactly equal to that in the white-light QPP.
Moreover, the quasi-period at the sunspot penumbra appears to be
always present and to not only exist in the time interval of the X6.4
flare. On the other hand, panels~(c) and (d) show the Morlet wavelet
power spectrum and its global wavelet power spectrum at the sunspot
umbra. They both reveal a quasi-period centered at about
3~minutes, which is much shorter than that at the sunspot penumbra
and white-light QPP. Similarly, the long running window of
15~minutes is also used for detrending, and their global wavelet
power spectra are shown in Figure~\ref{wav2}~(b) and (d). One can
note that the quasi-period of interest is still dominated in the
global power spectra, as indicated by the magenta dashed curves. The
wavelet analysis results also agree with the spatial distributions of
normalized Fourier power in Figure~\ref{pow}.

\begin{figure}
\centering
\includegraphics[width=\linewidth,clip=]{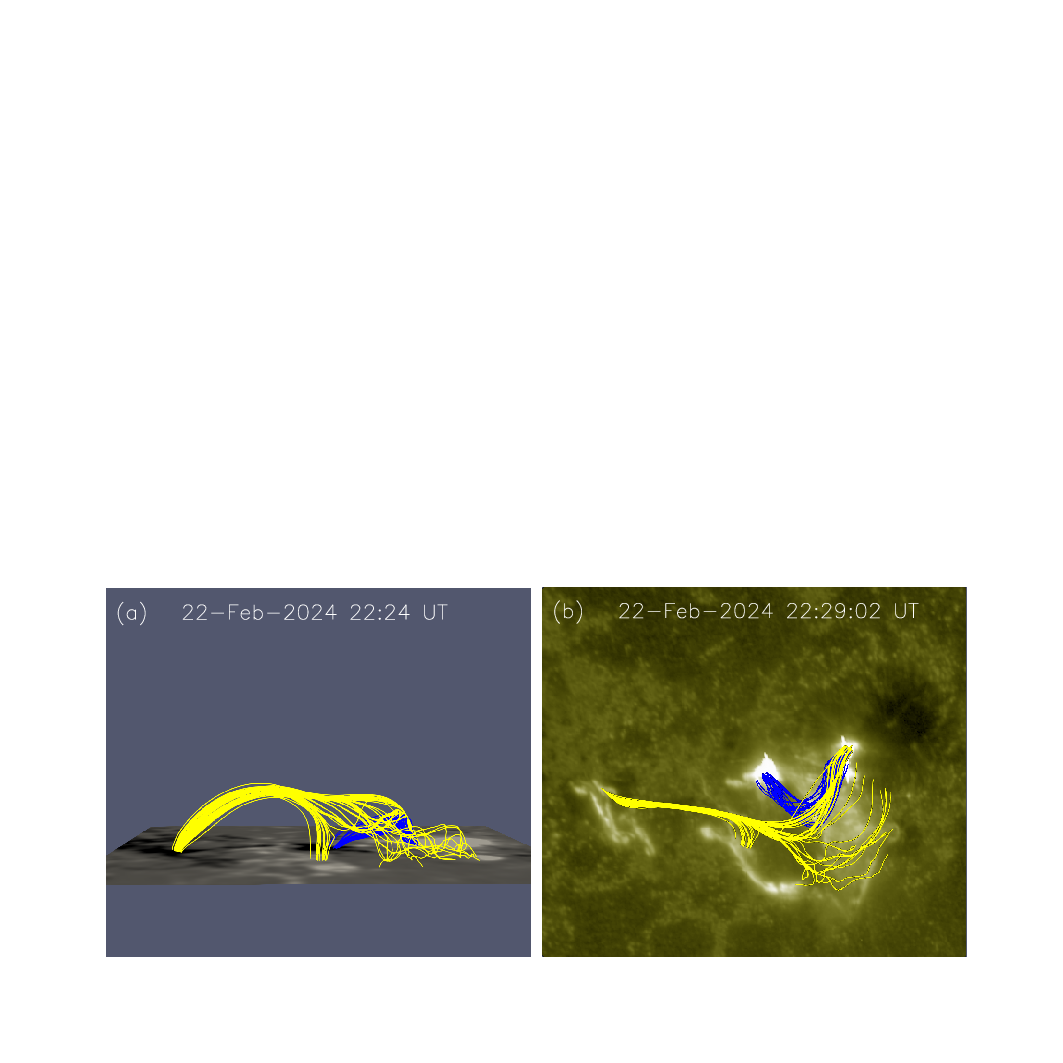}
\caption{Magnetic structures derived by the NLFFF extrapolation. (a)
Selected magnetic field lines overlaid on the HMI magnetogram,
viewed from the front side. (b) Selected magnetic field lines
overlaid on the AIA~1600~{\AA} map, viewed from the top. The yellow
lines represent a spine-fan topology, and the blue lines indicate
the overlying magnetic loops. \label{nlff}}
\end{figure}

Figure~\ref{nlff} shows the non-potential magnetic configuration for
the active region hosting the X6.4 flare derived by the nonlinear
force-free field (NLFFF) extrapolation. The NLFFF extrapolation was
performed by using the weighted optimization approach
\citep{Wheatland00} with the vector magnetograms measured by
SDO/HMI. Before the extrapolation, we rebinned the boundary data by
2$\times$2 to 0.72~Mm~pixel$^{-1}$. In this case, the extrapolation
region is a box of 566$\times$248$\times$248 with uniform grid
points, which corresponds to about 408$\times$179$\times$179~Mm$^3$.
For the extrapolated magnetic field, the current-weighted average of
the angle $<\theta_i>$ is smaller than 10$^{\circ}$. It should be
pointed out that the extrapolated magnetic fields satisfied both the
force-free and divergence-free conditions. We note that the main
quasi-circular profile and a remote ribbon-like feature seen in
UV/EUV maps are overlying on a spine-fan magnetic topology, as
indicated by the yellow lines, confirming the presence of the
circular-ribbon flare. The two patches of white-light brightening
are connected by a magnetic loop system that is rooted in the sunspot
penumbra, as indicated by the blue lines. The NLFFF-extrapolated
results suggest that the white-light QPP of the circular-ribbon
flare should be highly associated with the penumbral oscillation in
the nearby sunspot.

\section{Discussions}
Until now, the X6.4 flare is the second most powerful flare since the
beginning of the solar cycle~25, and the Sun becomes more active
than expected. The solar flare shows a significant enhancement in
the white-light continuum channel, which is simultaneously observed
by the ASO-S/WST in the white-light wave band of 3600~{\AA}, the
ASO-S/FMG in the optical continuum near the Fe 5324~{\AA} line, and
the SDO/HMI in the visible continuum near the Fe~6173~{\AA} line.
Thus, it could be regarded as a WLF. The WLF reveals four apparent
successive pulsations with a very long period in the white-light
wave band measured by HMI and WST. Next, a wavelet transform with the
mother function \citep{Torrence98} is used to determine the
quasi-period of the white-light QPP, which is about
8.6$^{+1.5}_{-1.9}$~minutes. The very long-periodic pulsations
(VLPs) have been observed in wavelengths of SXR and H$\alpha$,
respectively. However, those VLPs were observed in the pre-flare
phase and were considered as the prediction of powerful flares
\citep{Tan16,Li20a}. The flare QPPs with a quasi-period of roughly
8~minutes were simultaneously found in wave bands of the white-light
and UV radiation at the loop-top region during an X8.2 flare, which
was an off-limb WLF \citep{Zhao21}. In our case, the VLPs can be seen
from the impulsive phase to the decay phase of the X6.4 flare in the
white-light wave band, and such VLPs cannot be found in wavelengths
of UV and EUV. Moreover, the X6.4 flare manifested as two
brightness-enhancement patches on the solar disk, regarded as an
on-disk WLF, which is absolutely different from previous
observations \citep{Tan16,Li20a,Zhao21}.

Fare QPPs on the Sun are frequently observed in wavelengths of
SXR/HXR, UV/EUV, and radio/micowave
\citep{Dolla12,Dominique18,Hayes20,Lorincik22,Collier23,Karlicky23,Zimovets23,Lid24c,Millar24},
and they have been reported in the H$\alpha$
\citep{Li20a,Kashapova20}, Ly$\alpha$ \citep{Li21,Li22a}, and
$\gamma$-rays \citep{Nakariakov10,Li22b} emissions. However, the
flare QPPs in the white-light emission, named as white-light QPPs,
are rarely reported in solar flares \citep{Zimovets21}. In this
case, the white-light QPP is observed in the white-light emission at
WST~3600~{\AA} and the visible continuum radiation near the
Fe~6173~{\AA} line, but it cannot be seen in the UV emission. That
is, the flare QPP with a very long period can only exist in the
photosphere, and it cannot propagate upwardly into the chromosphere and
transition region on the solar disk. Thus, it could be regarded as
a pure white-light QPP. The modulation depth of the white-light
QPP, which is identified as the ratio of the oscillating amplitude
and its long-term trend, is quite small, i.e., $<$0.1\%, indicating
a weak signal of the white-light QPP. This is reasonable, because
the background of the optical light radiation in the photosphere is
rather strong, while the white-light enhancement of a solar flare is
much weaker than the background radiation, and thus the modulation
depth is very small compared to its background. This is also one
reason that why the white-light QPP is rarely reported on the Sun.

It is necessary to discuss the triggered mechanism of the
white-light QPP at a very long quasi-period. Generally, the flare
radiation in the white-light channel is strongly associated with
nonthermal electrons accelerated by the magnetic reconnection
\citep{Ding03,Li23}. However, the VLPs are only seen in the visible
continuum channel, and they cannot be observed in wavelengths of
microwave, UV/EUV, and SXR/HXR. Moreover, they can be found both in the
impulsive and decay phases during the X6.4 flare. Based on these
observational facts, the white-light QPP could not be triggered by
the quasi-periodic regime of magnetic reconnection, which is
definitely different from previous observations
\citep{Reeves20,Zhao21}. It is well accepted that the slow wave is a
weak wave process, and the white-light QPP studied here is also a
weak signal. Therefore, the white-light QPP at a very long period
might be modulated by the slow-mode MHD wave in the photosphere.
\cite{Sych15} have reported the possibility of flare
triggering by slow-mode waves propagating from a sunspot though a
selected magnetic loop to the flare site. We also note that the
X6.4 flare occurs near some sunspot groups, and the white-light
radiation tends to locate in the sunspot penumbra. As can be seen in
the visible continuum maps measured by HMI and FMG, both the two
bright patches of white-light enhancements appear in the penumbral
region, which are seated in two different sunspots. Thus, we
conjectured that the very-long period may be associated with the
penumbral oscillation at sunspots \citep{Loughhead58,Su13,Lid20}.
The very similar quasi-period of $\sim$8.5$^{+1.6}_{-1.8}$~minutes
is detected in the sunspot penumbral position, which is close to the
penumbral-photospheric boundary and far away from the
umbral-penumbral boundary. This is consistent with the previous
finding that the quasi-periods of penumbral oscillations become
longer and longer when they stay away from the umbra and approach to
the photosphere \citep{Yuan15,Su16,Feng20}. Similar to what has been
seen in the white-light QPP, the very long pulsations at the sunspot
penumbra can be only seen in the visible continuum channel.
We cannot find any apparent signature of the similar quasi-period in the
UV wave band of AIA~1600~{\AA} above the sunspot penumbra. This
observational feature suggests that the slow-mode MHD wave
originating from the sunspot penumbra cannot penetrate into the
solar upper atmosphere, such as the chromosphere, the transition
region, or even the corona on the Sun. This is consistent with the
theoretical expectation, that is, a magnetoacoustic cutoff
frequency is always existent above the sunspot, and only the
short-period wave can spread into the solar upper atmosphere
\citep{Sych09,Sych14,Yuan14}. Moreover, the penumbral oscillation at
the very long quasi-period appears to always exist, including the
time duration of nonflare and the flare eruption. Their modulation
depth (such as 2\%$-$3\%) is much larger than that of the
white-light QPP (i.e., $<$0.1\%). Thus, the penumbral oscillation is
basically a weak wave process, and it can influence and be a trigger
for the weaker white-light QPP. At last, the two bright patches in
white-light emissions are connected by magnetic loops that are
rooted in sunspot penumbras, which are demonstrated by the NLFFF
extrapolation. Based on those observational facts and the NLFFF-extrapolated results,
we can conclude that the white-light QPP at
the very long quasi-period is most likely to be modulated by a
slow-mode magnetoacoustic gravity (MAG) wave that is originating
from the sunspot penumbra.

\section{Summary}
Using the spaced-based instruments (ASO-S, GOES, and SDO), combined
with the ground-based telescopes (NoRP and EOVSA) and the NLFFF
extrapolation, we investigated the flare QPP and penumbral
oscillation in the white-light channel. Our main conclusions are
summarized as follows:

(1) The white-light QPP at a very long quasi-period is simultaneously
observed by SDO/HMI and ASO-S/WST. The flare QPP is manifested as
four successive VLPs in channels of the
HMI visible continuum and WST~3600~{\AA}. However, it cannot be
observed in wavelengths of HXR, microwave, and UV/EUV.

(2) A quasi-period of about 8.6$^{+1.5}_{-1.9}$ is detected in the
visible continuum flux. Moreover, the VLPs appear in both the
impulsive and decay phases, suggesting that they could not be
associated with the nonthermal electrons periodically accelerated by
magnetic reconnection. The modulation depth of the white-light QPP
is less than 0.1\%, suggesting a weak QPP signal.

(3) A quite approximate quasi-period of about
8.5$^{+1.6}_{-1.8}$~minutes is observed in the sunspot penumbra in
the photosphere, which is close to the penumbral-photospheric
boundary. The very long quasi-period can persist for an extended
period of time, but it cannot be found in the chromosphere,
transition region, and corona, indicating that it hardly penetrates
into the solar upper atmosphere.

(4) The white-light QPP is most likely to be modulated by the
slow-mode MAG wave that is originated from the sunspot penumbra. The
very long quasi-period is cut off in the photosphere, and only the
short-period wave can propagate upwardly to the solar upper
atmosphere.

\acknowledgments The authors would like to thank the referee for
his/her constructive comments. This work is supported by the
Strategic Priority Research Program of the Chinese Academy of
Sciences, grant No. XDB0560000, and the National Key R\&D Program of
China 2021YFA1600502 (2021YFA1600500), and 2022YFF0503002
(2022YFF0503000). D. Li is also supported by Yunnan Key Laboratory
of Solar Physics and Space Science under the number YNSPCC202207. We
thank the teams of ASO-S, SDO, GOES, NoRP, and EOVSA for their open
data use policy. The ASO-S mission is supported by the Strategic
Priority Research Program on Space Science, the Chinese Academy of
Sciences, grant No. XDA15320000.


\end{document}